\documentclass[useAMS,usenatbib]{mn2e}
\usepackage{graphicx,rotate}
\usepackage{natbib}
\usepackage{latexsym,amssymb,verbatim}

\title[Star formation indicators]{Are $^{12}$CO lines good indicators of the star formation
 rate in galaxies?}
\author[Bayet et al.]{E. Bayet$^{1}$\thanks{E-mail:
eb@star.ucl.ac.uk; gerin@@lra.ens.fr; tgp@submm.caltech.edu; contursi@mpe.mpg.de}; M. Gerin$^{2}$; T. G. Phillips$^{3}$ and
 A. Contursi$^{4}$\\
$^{1}$Department of Physics and Astronomy, University College
London, Gower Street, London WC1E 6BT, UK\\
$^{2}$LERMA, Observatoire de Paris and Ecole Normale Sup\'erieure, 24
rue Lhomond, F-75005 Paris, France (CNRS-UMR 8112)\\
$^{3}$California Institute of Technology, Downs Laboratory of Physics
 320-47, Pasadena, CA 91125, USA\\
$^{4}$Max Planck Institute f\"ur Extraterrestrische Physik, Postfach
 1312, 85741, Garching, Germany}

\begin{document}

\date{Accepted ; Received ; in original form }

\pagerange{\pageref{firstpage}--\pageref{lastpage}} \pubyear{2008}

\maketitle

\label{firstpage}

\begin{abstract}
In this paper, we investigate the relevance of using the $^{12}$CO
line emissions as indicators of star formation rates (SFR). For
the first time, we present this study for a relatively large
number of $^{12}$CO transitions (12) as well as over a large
interval in redshift (from z$\sim$0 to z$\sim$6). For the nearby
sources (D$\leq$10 Mpc), we have used homogeneous sample of
$^{12}$CO data provided by Bayet et al. (2004, 2006), mixing
observational and modelled line intensities. For higher-z sources
(z $\geq$ 1), we have collected $^{12}$CO observations from
various papers and have completed the data set of line intensities
with model predictions which we also present in this paper.
Finally, for increasing the statistics, we have included recent
$^{12}$CO(1-0) and $^{12}$CO(3-2) observations of intermediate-z
sources. Linear regressions have been calculated for identifying
the tightest SFR-$^{12}$CO line luminosity relationships. We show
that the \emph{total} $^{12}$CO, the $^{12}$CO(5-4), the
$^{12}$CO(6-5) and the $^{12}$CO(7-6) luminosities are the best
indicators of SFR (as measured by the far-infrared luminosity).
Comparisons with theoretical approaches from Krumholz and Thompson
(2007) and Narayanan et al. (2008) are also performed in this
paper. Although in general agreement, the predictions made by
these authors and the observational results we present here show
small and interesting discrepancies. In particular, the slope of
the linear regressions, for J$_{upper}\geq$ 4 $^{12}$CO lines are
not similar between theoretical studies and observations. On one
hand, a larger high-J $^{12}$CO data set of observations might
help to better agree with models, increasing the statistics. On
the other hand, theoretical studies extended to high redshift
sources might also reduce such discrepancies.
\end{abstract}


\section{Introduction}

Fifty years ago, it has been showed that the star formation rate
(hereafter SFR) is intimately linked with the reservoir of the gas
from which stars are forming \citep{Schi59}. The Kennicutt-Schmidt
power law parameterized on local galaxies by \citet{Kenn98a,
Kenn98b, Kenn07} has led to an SFR index of $N$=1.4$\pm$0.15.
Since late 90s, researchers have converted the Kennicutt-Schmidt
law into a more interesting relationship connecting the SFR
(traced by the infrared luminosity - hereafter L$_{ir}$) to the
mass of molecular gas, investigating various molecular tracers.
Firstly, \citet{Sand91, Sand96} have showed that the SFR is
roughly proportional to the $^{12}$CO(J=1-0) line (hereafter
CO(1-0)) luminosity (slope of 1.4-1.6, consistent with the
Kennicutt-Schmidt law index). This SFR-CO(1-0) relationship has
been broadly interpreted as an increase of star formation
efficiency as a function of molecular gas mass (and density). More
recent analysis focussing on tracers of \emph{dense} molecular gas
such as the HCN(1-0) line \citep{Gao04a, Gao04b, Wu05}, the
HCN(3-2) line \citep{Buss08} or the CO(3-2) line \citep{Yao03,
Nara05}, have shown that these transitions are likely to be better
indicators of star formation rate than the total H$_{2}$ content
(traced by the CO(1-0) luminosity). Indeed, the CO(1-0) line can
be excited at rather low densities ($\sim 10^{2}-
10^{3}$cm$^{-3}$) and low temperature ($\sim$ 5K above ground)
whereas the higher-J CO transitions and the HCN lines trace denser
and warmer gas, more closely connected with the stars in
formation. However, recent modelling work \citep{Krum07, Nara08}
show that this situation is more complex than it appears,
involving in particular the values of the molecular line critical
densities as compared with the mean gas density of the observed
regions.

So far, none of observational or modelling work investigate the
relationship between the SFR and molecular gas tracers both for
\emph{several transitions} ($>$2) of the same molecule (here CO)
and over \emph{several factors of redshift} (here z$\approx$0-6).
In this paper, the influence of the high-J CO lines is especially
studied. This is motivated by the fact that, in addition to obtain
much more accurate estimation of the CO line luminosities when
adding the high-J transitions (see \citealt{Baye04, Baye06}), more
subtle effect may be revealed by these lines, when considering
their respective critical densities.

The knowledge of the global star formation history, and for
individual galaxies, of their actual and past star formation rate,
is a key item for understanding galaxy evolution, and for
comparing with state-of-the-art models. Statistical relations
linking star formation properties with other galaxy global
characteristics established by observing nearby galaxies, have
been already used to understand the processes ruling the large
scale star forming activities (e.g. \citealt{Malh01, Bose02}).
Such trends can afterwards be used in the large scale cosmological
models, which lack the spatial and temporal resolution to describe
the local details of star formation in individual galaxies.

In this paper, we have thus compared the CO lines emissions (from
J=1-0 to J=12-11 transitions) with the infrared emission, for the
nearby galaxies we have surveyed using the Caltech Submillimeter
Observatory (CSO) (see \citealt{Baye04, Baye06}). We extended this
comparison to higher redshifted sources ($z = $ 1.4-6.4) for which
CO line emissions have been previously observed \citep{Cox02,
Bert03, Pety04, Walt04, Grev05, Solo05, Tacc06, Weis07}. For these
distant objects, we derived the \textit{total} and the individual
CO line emissions (from J=1-0 to J=12-11) by the same approach as
the one presented in \citet{Baye04, Baye06}. In order to
strengthen the results in a more statistical point of view, we
finally have included additional literature CO(1-0) and CO(3-2)
data from \citet{Gao04a, Gao04b, Yao03} and \citet{Nara05},
respectively.

The paper is divided straightforwardly as followed: Sect.
\ref{sec:sample} presents the data sample while Sect.
\ref{sec:resu} lists the results we have obtained, focussing
especially on the comparison with the model predictions from
\citet{Krum07} and \citet{Nara08}. Finally, we conclude in Sect.
\ref{sec:conclu}.

\section{Sample selection}\label{sec:sample}

It has been crucial to determine, or find in the literature, the
total CO line emission as well as the individual line emissions of
the CO transitions from J=1-0 to J=12-11. The total infrared
luminosity (from 8 to 1000 $\mu$m) for both the nearby and the
high-z sources has also been crucial to determine. Here, we have
converted all the gathered data, for the first time, into a
\textit{coherent} and \textit{homogeneous} sample we have
corrected for various effects, as described below. In Table
\ref{tab:lines} and Fig. \ref{fig:CO_curves}, the observed CO data
for nearby and distant sources are presented, respectively. Model
predictions go up to the J=15-14 transition of CO, consistently to
the work of \citet{Baye04, Baye06}. However, we have restricted
our study to only the first 12 CO transitions, higher-J CO lines
having very weak intensity (see \citealt{Baye04, Baye06} and Fig.
\ref{fig:CO_curves}). In the rest of the paper, the \emph{total}
CO will thus refer to the sum of only the first 12 CO transitions.

\begin{table}
\caption{List of the detected CO lines for nearby galaxies
included in this study. These CO lines detections are from
\citet{Baye04, Baye06}. Similar information for distant sources is
found in Fig. \ref{fig:CO_curves} (observations represented by
black bullets). When the line is not detected, we have used its
corresponding predicted emission from modelling work (see
Subsects. \ref{subsec:nearby} and \ref{subsec:highz}). For the
Antennae sources (NGC 4038 and Overlap), the $^{12}$CO(4-3) and
$^{12}$CO(7-6) lines have been observed by \citet{Baye06} but they
suffer from large uncertainties. Therefore, we used for these two
objects and two transitions, rather the corresponding best model
predictions.} \label{tab:lines}
\begin{tabular}{lcll}
\hline
Name & Distance & detected $^{12}$CO lines\\
\hline
\textbf{Nearby sources} & (Mpc) &\\
IC 10 & 1.0$^{1}$ & 1-0, 2-1, 3-2, 4-3, 6-5, 7-6\\
NGC 253 & 2.5$^{2}$ & 1-0, 2-1, 3-2, 4-3, 6-5, 7-6\\
IC 342 & 1.8$^{3}$ & 1-0, 2-1, 3-2, 4-3, 6-5, 7-6\\
He 2-10 & 9.0$^{4}$ & 1-0, 2-1, 3-2, 4-3, 6-5, 7-6\\
NGC 4038 & 13.8$^{5}$ & 1-0, 2-1, 3-2, 4-3, 6-5, 7-6\\
Overlap$^{a}$ & 13.8$^{5}$ & 1-0, 2-1, 3-2, 4-3, 6-5, 7-6\\
M 83 & 3.5$^{6}$ & 1-0, 2-1, 3-2, 4-3, 6-5\\
NGC 6946 & 5.5$^{7}$ & 1-0, 2-1, 3-2, 4-3, 6-5\\
\hline
\end{tabular}

$^{a}$: Overlap corresponds to a shifted position from the NGC~4039
nucleus in the Antennae Galaxy which shows a high gas density (see
\citealt{Baye06}). References: $^{1}$ : \citet{Mass95}; $^{2}$ :
Adopted value from \citet{Maue96}; $^{3}$ : \citet{McCa89}; $^{4}$ :
\citet{Kobu99}; $^{5}$ : \citet{Savi04}; $^{6}$ : \citet{Thim03} and
$^{7}$ : \citet{Tull88}.
\end{table}

\subsection{Nearby sources}\label{subsec:nearby}

The CO data we are using in this study for the nearby sources are
from the observational and modelling work of \citet{Baye04,
Baye06}. They already provided a consistent line intensity sample
for 8 nearby ($<$10 Mpc) galaxies: NGC~253, IC~10, IC~342,
NGC~6946, M~83, Henize~2-10 and the Antennae (NGC~4038 and
Overlap) from CO(1-0) to CO(15-14). We converted the first twelve
CO integrated line intensities into luminosities (in
Kkms$^{-1}$pc$^{2}$) using the formulae in \citet{Solo05}:

\begin{equation}\label{eq:lumi}
L'_{CO} = 3.25 \times 10^{7} \times S_{CO}\Delta v \times \nu_{obs}^{-2}
\times (1+z)^{-3} \times D_{L}^{2}
\end{equation}

where z is the redshift\footnote{For nearby sources we used the
NED database values of z.}, $\nu_{obs}$ is the observed
frequency\footnote{The observed frequency is equal to the rest
frequency $\nu_{rest}$ divided by (1+z).} (in GHz) of the CO line,
$S_{CO}\Delta v$ is the CO integrated line intensities (in Jykms$^{-1}$)
and $D_{L}$ is the luminosity distance (in Mpc)\footnote{For all the
sources, we obtained the luminosity distance using the web site
calculator of
http://www.astro.ucla.edu/$\sim$wright/CosmoCalc.html, within a
cosmology of H$_{o}=$ 77 kms$^{-1}$Mpc$^{-1}$, $\Omega_{M}=$0.27
and $\Omega_{V}=$0.73.}. The total CO luminosity has been obtained by summing the individual CO line luminosities converted previously from Eq. \ref{eq:lumi} into solar luminosity units (L$_{\odot}$) to avoid any frequency bias in the SFR-total CO relationship. Table \ref{tab:data} summaries the
results obtained.

The infrared data used for the nearby sources and, more generally
in the whole paper, are from \citet{Sand03}. For consistency with
the additional literature data described in Subsect.
\ref{subsec:lite}, we chose to use the total infrared luminosity
from 8 $\mu$m to 1000 $\mu$m (L$_{IR:8-1000\mu m}$), thereby
including the continuum emission shorter than 60 $\mu$m affected
by the contribution from very small grains. This total infrared
luminosity is known to be contaminated by a possible AGN
contribution (see also \citealt{Grac08}). In particular, the AGN
yield can be fairly large in the MIR range (e.g \citealt{Rowa89})
in active galaxies such as LIRGs and ULIRGs. However, for the
sample of the nearby sources we are studying here, this AGN
contamination is considered as negligible since none of the 8
sources (IC~10, NGC~253, IC~342, Henize~2-10, NGC~4038, Overlap,
M~83 and NGC~6946) is known for hosting an AGN (see e.g.
\citealt{Lero06, Mart06, User06, Wils00, Kram05, Isra01},
respectively).

A more important aspect for nearby sources is that the CO
measurements (both observations and model predictions) referred to
an aperture of 22$''$, as described by \citet{Baye04, Baye06},
while the infrared data from \citet{Sand03} showed a higher
aperture (80$''$). Thus, in Fig. \ref{fig:corre} where we present
various SFR-CO luminosity relationships (see Sect.
\ref{sec:resu}), we compare, for nearby sources, two emitting
regions which are not spatially identical. Indeed, the region
which emits the infrared luminosity is larger than the region
where CO is detected. It is the same case for the intermediate-z
data (Subsect. \ref{subsec:lite}) but this problem does not appear
in the case of higher-z sources since they are seen as point-like
sources in both the sub-mm/mm and infrared wavelengths. To correct
this effect on nearby and intermediate-z sources, we have
estimated a factor between the CO and the infrared luminosities
resolution via the dust emission traced by SCUBA maps at 850
$\mu$m. We have derived from these maps the luminosity (removing
the background contribution) of the 850 $\mu$m emission at a
spatial resolution of 22$''$ and 80$''$, for the sources available
in the SCUBA archive. We have then calculated the ratio between
the emissions observed at 22$''$ and 80$''$ that we have applied
to the infrared data. We have obtained ratios varying from 2.3 to
6.5, depending on the source. Rather than the 450 $\mu$m SCUBA
maps, we have chosen the 850$\mu$m SCUBA maps because they show a
better signal-to-noise ratio. Nevertheless, we checked that the
factors obtained at 850$\mu$m were in agreement with the ones at
450$\mu$m (difference obtained being less than 2\%). After having
applied such correction to the infrared data of the galaxies in
common to this paper and the SCUBA archive, we have however
obtained similar results to those listed in Table
\ref{tab:resu_regres} and presented in Fig. \ref{fig:corre} which
show non-aperture corrected data. The slopes did not show changes
greater than 5\% in their values, depending on the SFR-CO line
luminosity relationship studied. Due to the fact that all the data
could not be corrected consistently for this effect since the
galaxies presented here have not been all observed by SCUBA (e.g
IC~342), we thus have decided to keep non-corrected infrared data
in Fig. \ref{fig:corre}, increasing however the error bars on the
slope values by 5\% in Table \ref{tab:resu_regres}.

Both the (non-aperture corrected) infrared and the CO luminosities used in
the Fig. \ref{fig:corre} are listed in Table \ref{tab:data}.

\subsection{High-z sources}\label{subsec:highz}

So far, a complete \emph{observed} CO SED does not exist for
J=1-0 to J=12-11 neither for nearby nor high-z sources.

At high redshift, various CO transitions have been however already
detected (e.g. \citealt{Cox02, Bert03, Pety04, Walt04, Grev05,
Solo05, Tacc06, Weis07}). Most of these data are summarized in
\citet{Solo05}. In our study, we restricted the number of studied
sources presented in \citet{Solo05} to ten objects : 4C60.07, APM
08279\footnote{For this source, we rather used more recent CO data
from \citet{Weis07}}, Cloverleaf QSO, SMM J14011, VCV J1409,
PSSJ2322, TN J0924, SDSS J1148, IRAS F10214 and HR 10, keeping
only those which show at least two CO line detections. Indeed, to
correctly constrain the models which estimate the missing line
emissions (LVG models, see Appendix A) and derive relevant
predicted CO line intensities, it is crucial to have as many CO
line detections as possible. This is why we have rejected from our
study high-z sources presented in \citet{Solo05} with only one CO
line observed.

\begin{figure*}
    \centering
    \includegraphics[width=9cm]{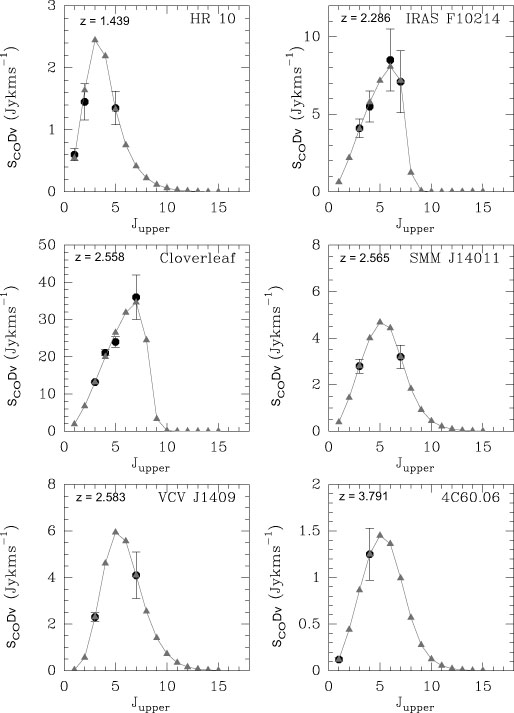}
    \includegraphics[width=9cm]{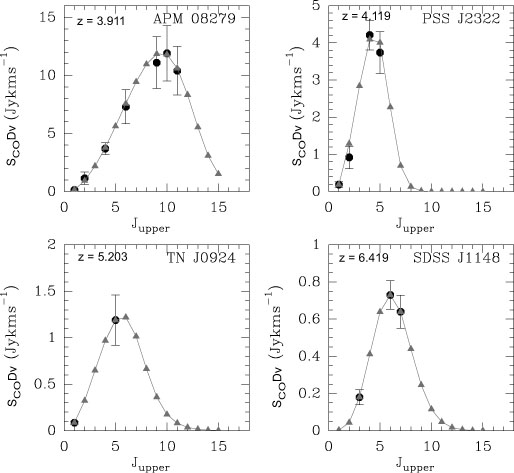}
    \caption{Observed and predicted CO spectral energy distributions
    (Jykms$^{-1}$) of the following high redshift sources (from top to bottom, and by
    increasing order of redshift):
    HR 10, IRAS F10214, The Cloverleaf QSO, SMM J14011, VCV J1409,
    4C60.06, APM 08279, PSS J2322, TN J0924 and SDSS J1148 (see plots).
    Observations and their corresponding error bars are represented
    by black bullets (see references listed in \citealt{Solo05}) while
    LVG predictions (see Appendix A) are symbolized with grey
    triangles. To make the plots more easily
    readable, we have connected
    the CO LVG predictions (grey lines). One could notice that, except
    for the source HR~10, the position of the maximum of the SED
    is located at rather high-J (J$_{upper}\geq$4).}\label{fig:CO_curves}
\end{figure*}

\begin{figure*}
    \centering
    \includegraphics[width=17cm]{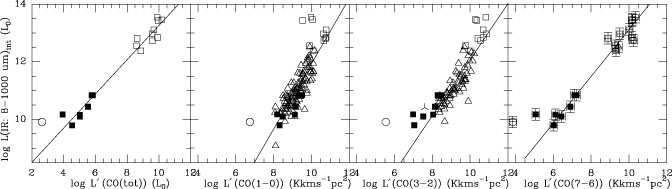}
    \caption{Examples of SFR-CO line luminosity
    relationships we have obtained,
    expressed in log-log scales. From left to right, we plotted the
    total integrated infrared luminosity L$_{IR:8-1000\mu m}$ (in
    L$_{\odot}$) versus the total CO (in
    L$_{\odot}$), the CO(1-0), the CO(3-2) and the
    CO(7-6) luminosities (in Kkms$^{-1}$pc$^{2}$) (see
    Sect. \ref{sec:sample}). The squares correspond to the sources we
    studied in details in this paper (see
    Subsects. \ref{subsec:nearby} and \ref{subsec:highz}) while the
    triangles correspond to the literature data we have added from
    \citet{Yao03, Gao04a, Gao04b, Nara05} (see
    Subsect. \ref{subsec:lite}). The black filled squares represent
    the nearby sources while the white opened squares symbolize the
    high-z sources. In all plots, IC~10 is separated from
    other nearby sources (see Subsect. \ref{subsec:gene}) and
    represented by open white circles. In the plot representing the
    SFR-CO(3-2) luminosity relationship one can see a lambda symbol
    corresponding to the source NGC~7817 we have also excluded (see Subsect.
    \ref{subsec:gene}). The linear regressions
    have been obtained using the xmgrace software. They
    are represented by black lines. For keeping the figures clear, we
    plotted the typical errors bars of the observations (included in the
    calculations of the linear regressions), only on the
    SFR-CO(7-6) luminosity correlation. These errors
    correspond to a $\pm 1\sigma$ uncertainties both on the
    L$_{IR}$ and on the molecular CO line luminosities.
    }\label{fig:corre}
\end{figure*}

To estimate the \textit{total} CO luminosity as well as the
individual missing molecular CO line luminosities in these
sources, we have used a single-component Large Velocity Gradient
(LVG) model as done in \citet{Baye04, Baye06}. The obtained line
intensity predictions (see Fig. \ref{fig:CO_curves}) correspond to
a beam size of 22$''$, consistently with other data sets. The goal
of the paper is to investigate the SFR-molecular CO line
luminosity relationships. Thus, we will not discuss further the
physical properties of the molecular gas we have obtained using
the LVG model. Nonetheless, we present them in Appendix A.

In Fig. \ref{fig:CO_curves}, we have superimposed on the CO
observations (black bullets with error bars), the predicted
emissions of both the observed and the missing CO lines (grey
filled triangles). Similarly to the nearby sources case, the total
CO luminosity has been obtained by summing the individual CO line
luminosities converted previously from Eq. \ref{eq:lumi} into
solar luminosity units (L$_{\odot}$) to avoid any frequency bias
in the SFR-total CO relationship. More precisely, we have used the
observed velocity-integrated CO line fluxes ($S_{CO}\Delta v$ in
Jykms$^{-1}$) when detected and the predicted values when not. The
redshift values used in Eq. \ref{eq:lumi} are those presented in
\citet{Solo05}.

We did not need to correct the CO line luminosities for any beam
dilution effect since the high-z galaxies are point-like sources
within telescope beams and show unresolved emissions whatever
these wavelengths. However, some of these distant objects are
lensed. Thus, we have applied the factor of lens magnification
listed in \citet{Solo05} to \textit{all} the CO line luminosities.
The underlying assumption of such a correction is that the
gravitational lens magnifies similarly the emission from compact
and warm regions usually traced by high-J CO lines such as the
CO(7-6) line and the regions more extended normally traced by
typical low-J lines such as the CO(1-0) transition. In principle,
this is a relevant assumption since the properties of the lens
does not depend on the magnified source but only on the properties
of the galaxies separated the observer from the studied source.

For the infrared luminosity, we used the integrated values
(``L$_{FIR}$(int.)'') listed in Table 1 of \citet{Solo05},
corresponding to infrared emission corrected for lens
magnification. After having checked several references in this
table, it appears that the ``L$_{FIR}$(int.)'' values are similar
to the L$_{IR:8-1000\mu m}$ presented in \citet{Sand03} (see the
example of HR~10 for which \citealt{Dey99} derived the infrared
luminosity by fitting the dust SED on a rest wavelength range from
10 $\mu$m to 2 cm.).

For these distant sources, we also present in Table
\ref{tab:data}, both the infrared and the CO luminosities we have
used in Fig. \ref{fig:corre}.

\subsection{Intermediate-z data from the literature}\label{subsec:lite}

We have included in our study CO data from observations at
intermediate redshift (within similar beam sizes) to better
constrain the SFR-molecular CO line luminosity relationships, and
especially increase the statistics. These data are either from
\citet{Gao04a, Gao04b} (CO(1-0) transition observed with a 22$''$
beam size), or from \citet{Yao03} and \citet{Nara05} (CO(3-2) line
observed with a 15$''$ and 22$''$ beam sizes, respectively). We
have not included the CO(2-1) data reported in \citet{Rigo96}
(beam size of 23$''$) because they appear inconsistent with those
presented in \citet{Yao03, Gao04a, Gao04b, Nara05}. We have
included in our SFR-CO relationships the observed and modelling
results dedicated to Markarian 231 \citep{Papa07}, providing CO
line luminosities estimations up to the CO(10-9). For the infrared
luminosity values, we have used the values reported in
\citet{Sand03} and not those listed in each corresponding papers.

The additional sources included here are seen to be mostly LIRGs
and ULIRGs (at a distance $>$ 10 Mpc). Radio continuum maps
\citep{Cond90}, and the HI and SCUBA 850 $\mu$m maps
\citep{Thom02}, show that much of the emission of the objects is
extended with respect to the beam size of 15$''$. However, in
recent high-resolution (2$''$-3$''$) millimeter wave observations
of seven LIRGs/ULIRGs, \citet{Brya99} find that nearly all of the
detected CO(J=1-0) emission is concentrated within the central
1.6kpc in six of seven objects. Because it is unlikely that there
will be significant high-J CO emission where there is no CO(1-0) detected,
we assume that all of the emitting gas in these additional sources
is confined within the same region. Consequently, we have not
converted the 14$''$ beam data into that for a 22$''$ beam.

As for nearby galaxies, the infrared luminosities of these sources
are expressed for a larger aperture than the one used for the CO
data. As previously mentioned (see Subsect. \ref{subsec:nearby}),
this effect on the SFR-CO line luminosity relationship has been
investigated through SCUBA data. Unfortunately, as for the case of
nearby galaxies, the intermediate-z sources we are using here have
indeed not been \textit{all} observed by SCUBA, therefore no
correction on their infrared data has been performed directly. We
have however taken fully into account the effect of different
spatial resolutions on the slope values listed in Table
\ref{tab:resu_regres} (increased error bars as explained in
Subsect. \ref{subsec:nearby}).

In the samples presented in \citet{Yao03, Gao04a, Gao04b, Nara05},
some sources such as ARP 220 are in common. For removing any
calibration effects on the SFR-CO relationships, we have followed
the recommendation of \citet{Nara05} and applied their scaling
factor equal to 0.26 for
the CO(3-2) and to 0.45 for the CO(1-0). They specified that this
effect may indeed occur when various instruments (and thus various
calibration processes) are used.

Note that the intermediate-z sources increase significantly the
number of CO observations used in the correlations either in the
SFR-CO(1-0) line luminosity or in the SFR-CO(3-2) line luminosity
relationships (see triangles in Fig. \ref{fig:corre} as well as
the fourth column in Table \ref{tab:resu_regres}). This
strengthens the corresponding results.

\begin{table*}
\caption{Infrared and CO lines luminosities of the sources in our
sample (see squares symbols in Fig. \ref{fig:corre}). For distant
galaxies, we also list in the last column the lens magnification
factor we have found in \citet{Solo05} and use in this study (see 
Fig. \ref{fig:corre}).
Intermediate-z sources values are not shown in this table since
already described in details in the references mentioned in the
text. When an asterisk is seen, it means that the value has been
estimated from observations only.}\label{tab:data}
\begin{tabular}{lcccccccccc} \hline
Name & L$_{IR}^{a}$ & L'$_{CO(tot)}^{b}$ &  Ref. & L'$_{CO(1-0)}$ &  Ref. &L'$_{CO(3-2)}$ &  Ref. &L'$_{CO(7-6)}$ &  Ref. & \\
     & L$_{\odot}$ & L$_{\odot}$ & & Kkms$^{-1}$pc$^{2}$ & & Kkms$^{-1}$pc$^{2}$ & & Kkms$^{-1}$pc$^{2}$ & & \\
\hline
IC 10  & 8.2$\times$10$^{9}$ & 4.4$\times$10$^{2}$ & 1 \& 2 & 5.6$\times$10$^{6*}$ & 1 \& 2 & 3.5$\times$10$^{5*}$ & 1 \& 2 & 7.3$\times$10$^{3*}$ & 1 \& 2 &\\
NGC 253 & 2.8$\times$10$^{10}$ & 3.3$\times$10$^{5}$ & 2 \& 3 & 1.5$\times$10$^{9*}$ & 1 \& 2 & 1.4$\times$10$^{8*}$& 1 \& 2 & 8.0$\times$10$^{6*}$ & 1 \& 2 &\\
IC 342 & 1.5$\times$10$^{10}$ & 8.4$\times$10$^{3}$ & 1 \& 2 & 1.6$\times$10$^{8*}$ & 1 \& 2 & 9.7$\times$10$^{6*}$& 1 \& 2 & 1.2$\times$10$^{5*}$& 1 \& 2 &\\
He 2-10 & 6.2$\times$10$^{9}$ & 3.4$\times$10$^{4}$ & 2 \& 3 & 2.1$\times$10$^{8*}$ & 1 \& 2 & 1.1$\times$10$^{7*}$ & 1 \& 2 & 1.0$\times$10$^{6*}$ & 1 \& 2 &\\
NGC 4038 & 6.9$\times$10$^{10}$ & 5.5$\times$10$^{5}$ & 1 \& 2 & 2.1$\times$10$^{9*}$ & 1 \& 2 & 1.9$\times$10$^{8*}$ & 1 \& 2 & 1.3$\times$10$^{7*}$ & 1 \& 2 &\\
Overlap & 6.9$\times$10$^{10}$ & 6.7$\times$10$^{5}$ & 1 \& 2 & 3.3$\times$10$^{9*}$ & 1 \& 2 & 2.9$\times$10$^{8*}$ & 1 \& 2 & 1.7$\times$10$^{7*}$ & 1 \& 2 &\\
M 83 & 1.3$\times$10$^{10}$ & 1.0$\times$10$^{5}$ & 1 \& 2 & 3.1$\times$10$^{8*}$ & 1 \& 2 & 3.3$\times$10$^{7*}$ & 1 \& 2 & 2.7$\times$10$^{6}$ & 1 \& 2 &\\
NGC 6946 & 1.5$\times$10$^{10}$ & 1.0$\times$10$^{5}$ & 1 \& 2 & 1.4$\times$10$^{9*}$ & 1 \& 2 & 1.1$\times$10$^{8*}$ & 1 \& 2 & 1.3$\times$10$^{6}$ & 1 \& 2 &\\
\hline
Name& L$_{IR}^{a}$ & L'$_{CO(tot)}^{b}$ &  Ref. & L'$_{CO(1-0)}$ &  Ref. &L'$_{CO(3-2)}$ &  Ref. &L'$_{CO(7-6)}$ &  Ref. & Lens\\
     & L$_{\odot}$ & L$_{\odot}$ & & Kkms$^{-1}$pc$^{2}$ & & Kkms$^{-1}$pc$^{2}$ & & Kkms$^{-1}$pc$^{2}$ & & magn. factor\\
\hline
HR10  & 6.5$\times$10$^{12}$ & 4.2$\times$10$^{8}$ & 2 & 5.6$\times$10$^{10*}$ & 4 \& 2 & 2.5$\times$10$^{10}$ & 2 & 7.9$\times$10$^{8}$ & 2 &1\\
IRAS F10214  & 3.6$\times$10$^{12}$ & 6.6$\times$10$^{9}$ & 2 & 1.4$\times$10$^{11}$ & 2 & 9.9$\times$10$^{10*}$ & 4 \& 2 & 3.2$\times$10$^{10*}$ & 4 \& 2 &17\\
Cloverleaf  & 5.4$\times$10$^{12}$ & 4.3$\times$10$^{10}$ & 2 & 5.0$\times$10$^{11}$ & 2 & 3.9$\times$10$^{11*}$ & 4 \& 2 & 1.9$\times$10$^{11*}$ & 4 \& 2 &11\\
SMM J14011  & 2.4$\times$10$^{12}$ & 6.1$\times$10$^{9}$ & 2 & 1.0$\times$10$^{11}$ & 2 & 8.3$\times$10$^{10*}$ & 4 \& 2 & 1.7$\times$10$^{10*}$ & 4 \& 2 &8.3\\
VCV J1409  & 3.5$\times$10$^{13}$& 7.8$\times$10$^{9}$ & 2 & 9.6$\times$10$^{9}$ & 2 & 6.9$\times$10$^{10}$ & 4 \& 2 & 2.3$\times$10$^{10}$ & 4 \& 2 &1\\
4CO60.07  & 1.3$\times$10$^{13}$ & 4.8$\times$10$^{9}$ & 2 & 6.1$\times$10$^{10*}$ & 4 \& 2 & 4.9$\times$10$^{10}$ & 2 & 1.0$\times$10$^{10}$ & 2 & 1\\
APM 08279  & 2.9$\times$10$^{13}$ & 1.1$\times$10$^{11}$ & 2 & 8.0$\times$10$^{10*}$ & 5 \& 2 & 1.3$\times$10$^{11}$ & 5 \& 2 & 1.0$\times$10$^{11}$ & 5 \& 2 &7\\
PSS J2322  & 9.3$\times$10$^{12}$ & 9.6$\times$10$^{9}$ & 2 & 1.1$\times$10$^{11*}$ & 4 \& 2 & 1.8$\times$10$^{11}$ & 2 & 8.3$\times$10$^{9}$ & 2 &2.5\\
TN J0924  & 7.0$\times$10$^{12}$ & 9.9$\times$10$^{9}$ & 2 & 7.3$\times$10$^{10*}$ & 4 \& 2 & 6.0$\times$10$^{10}$ & 2 & 1.7$\times$10$^{10}$ & 2 &1\\
SDSS J1148  & 2.7$\times$10$^{13}$& 9.1$\times$10$^{9}$ & 2 & 3.5$\times$10$^{9}$ & 2 & 2.2$\times$10$^{10}$ & 4 \& 2 & 1.5$\times$10$^{10}$ & 4 \& 2&1\\
\hline
\end{tabular}
$^{a}$ : L$_{IR}$ is the 8 -- 1000 $\mu$m total infrared
luminosity deduced from either \citet{Sand03} (nearby sources) or
\citet{Solo05} (high-z galaxies). For the distant objects,
L$_{IR}$ are corrected from the lens magnification effect (see
text in Subsect. \ref{subsec:highz}); $^{b}$ : L'$_{CO(tot)}$ is
the total CO line luminosity obtained after the use of models for
predicting the emission of the missing (non-observed) CO lines
(see the Subsects. \ref{subsec:nearby} and \ref{subsec:highz} and,
for more details, the Appendix A and the modelling work of
\citealt{Baye04, Baye06}). References : 1: See \citet{Baye06}; 2:
This work; 3: \citet{Baye04}; 4: \citet{Solo05}; 5: \citet{Weis07}
\end{table*}

\section{Results}\label{sec:resu}

\subsection{General arguments}\label{subsec:gene}

We present in Fig. \ref{fig:corre}, the SFR-CO line luminosity
relationships we have obtained for the total CO line emission, the
CO(1-0), CO(3-2) and CO(7-6) line emissions. In the same vein, we
have obtained the SFR-CO line luminosity relationships for the
other CO transitions: J=2-1 and from J=4-3 to J=12-11. SFR-CO
luminosity relationships for CO data higher than CO(7-6) are less
relevant than the correlation involving lower-J CO lines. Indeed,
most of the data in CO(8-7), CO(9-8), CO(10-9), CO(11-10) and
CO(12-11) are from modelled fit, similarly to the fits presented
in Fig. \ref{fig:CO_curves}. In such cases, correlations between
the SFR and the CO luminosity are not constrained by many
observational measures. Thus, we exclude these correlations from
the following analysis.

As one can see in Fig. \ref{fig:corre}, a
proportionality exists between the CO and infrared luminosities over a
large range of redshift (galaxies from z$\sim$ 0 up to z$\sim$ 6
sources). This characteristic has been previously seen
(\citealt{Sand91, Sand96} and references therein) although this
sample was more restricted in redshift than the one we
present in this paper. It has been interpreted as
an increasing star formation efficiency (SFE; SFR divided by
M(H$_{2}$)) as a function of molecular gas mass. Here, our data
sample confirms this broad interpretation.

To quantify better these relationships, we performed linear
regressions (using the software xmgrace). The corresponding output
parameters are listed in Table \ref{tab:resu_regres}. Note that
IC~10 has been excluded from these calculations. For the
SFR-CO(3-2) relationship we have also excluded NGC~7817, as
\citet{Nara05} suggested. Indeed these two sources show very small
L'$_{CO}$ value with respect to the locus of other objects. This is
especially true for IC~10 over several CO transitions (see Fig.
\ref{fig:corre}).

Deviations from the SFR-CO trend can be expected for subsolar
metallicity such as IC~10 (12 log(O/H)= 8.31 $\pm$ 0.2 from
\citealt{Arim96}). As shown by \cite{Lequ94}, a CO deficit (and a
[CII] excess), relatively to the infrared luminosity is expected
for low metallicity systems. At low metallicity, molecular
hydrogen can still be formed on grains, although the threshold
between atomic and molecular hydrogen is slightly shifted compared
to what is seen in our Galaxy (see the discussion on the FUSE
results by \citealt{Tuml02}). In addition, CO is more easily
destroyed by FUV photons than H$_2$. In diffuse and translucent
clouds, the decrease of CO photodissociation with increasing A$_v$
is provided by the combined effect of self-shielding and dust
extinction. At lower metallicities, these two effects get weaker,
hence CO can not survive at small A$_v$. In molecular clouds, CO
survives in the core, but an extended envelope where carbon is
ionized is also present. Therefore, the CO emission is decreased,
(and the [CII] emission increased) relative to normal metallicity
conditions.

Over the last years, in the literature, other tracers of molecular
gas have been investigated as SFR indicators, such as the HCN
\citep{Gao04a, Gao04b, Vandenbout04, Cari05, Buss08}. The linear
and tighter relationship shown between the HCN(1-0) line
luminosity and the infrared luminosity, as compared to the
relation SFR-CO(1-0), has been first interpreted by the fact that
the HCN is a molecule tracing warmer and denser gas, more closely
link with the stars in formation. Indeed, the CO(1-0) tends to
emit from both dense core and more diffuse molecular filaments and
cloud atmosphere (n$_{crit} \sim 10^{2}-10^{3}$ cm$^{-3}$) leading
to a non-linear relationship, whereas the HCN is typically only
thermalized in the dense cores of molecular clouds (n$_{crit} \sim
10^{5}$ cm$^{-3}$), allowing thus linearity. However, the fact
that, in a similar sample of galaxies \citet{Yao03, Nara05} also
found a linear SFR-CO(3-2) relationship provide evidence against
the sole HCN-related chemistry explanation (The critical density
of the CO(3-2) is n$_{crit} \sim 10^{4}$ cm$^{-3}$) .

More recent theoretical analysis \citep{Krum07, Nara08} have shown
that the relationship between the SFR and molecular line
luminosity in star-forming clouds (and thus, the value of the
slope of the linear regressions) depends rather on how the
critical density of the molecular transition compares to the mean
density of the observed sources. Lines with critical densities
smaller than the mean density in a observed region
(e.g. CO(1-0)) probe the total molecular gas and the SFR-molecular
line relationship slope is in agreement with the Kennicutt-Schmidt
index (slope $\sim$ 1.5). On the contrary, lines with critical
densities larger than the mean density (e.g. HCN(1-0) and CO(3-2))
show luminosities rising faster than linearly with increasing mean
gas density and the SFR-molecular line relationship is sub-linear
(slope $<$ 1.0). In this paper, we \emph{confirm observationally
these conclusions} (see Table. \ref{tab:resu_regres}) extending
them, in addition, to a larger interval in redshift.

\subsection{More detailed analysis}\label{subsec:deta}

Fig. \ref{fig:corre} and Table \ref{tab:resu_regres} show that the
SFR -CO(5-4),-CO(6-5) and -CO(7-6) line luminosity relationships
are the tightest correlations (highest values of the correlation
coefficients). Thus, between all the CO transitions, for the data
sample we present here, these transitions might be considered as
the \emph{best star formation rate indicators in galaxies}. The
total CO luminosity is also a very good tracer of star formation
rate in galaxies since it shows a tight correlation (correlation
coefficient of 0.984).

These tight SFR-high-J CO correlations contrast with the one seen
for the CO(1-0) line. Indeed, even if we restrict the sample of
CO(1-0) observations to only the 17 sources we used in the
SFR-higher-J CO line luminosity relationships, considering like
that the same statistics, the correlation coefficient of the
CO(1-0) line is still the lowest (see Table
\ref{tab:resu_regres}). This has been previously mentioned by
\citet{Riec06}. In the same vein, it has been shown by
\citet{Buss08} that the higher-J HCN line such as the HCN(3-2)
transition is more tightly correlated to the SFR traced by the
infrared luminosity than the HCN(1-0) line. We could thus conclude
that \emph{the CO(1-0) line is not a good indicator of star
formation rates in galaxies}.

In a more detailed analysis of the plots seen in Fig.
\ref{fig:corre}, it is interesting to note that, even if the
linear regressions fit rather well in all cases the observations,
they however become slightly not so relevant at low CO luminosity,
and thus, especially for the SFR-high-J CO relationships. This
tendency is even more obvious in the SFR-CO(3-2) plot. The
galaxies IC~342 and Henize~2-10 seem to be mainly responsible of
this divergence. In the SFR-CO(7-6) plot as well as in the
correlations involving higher-J CO lines, it is more difficult to
see this trend, probably because the statistics is reduced to 17
sources. This leads to a less accurate linear regression than in
the SFR-CO(3-2) case. In the SFR-CO(7-6) line luminosity
relationship, one could also notice that the deficit of CO
luminosity of the galaxy IC~342 is even more enhanced. The fact
that IC~342 is a less active
galaxy and that Henize~2-10 is small in our nearby source sample may also play a role. Indeed,
as compared to more active galaxies (such as NGC~253), they may
contain less dense gas. This has been recently confirmed by
\citet{Baye09} who detected the CS(7-6) line with much more
difficulties in IC~342 than NGC~253 (signal-to-noise ratio lower
by a factor of 3 between the two sources). Consequently, more
observations of high-J CO lines in quiet nearby galaxies are
needed to explain better this characteristic. Another possible
explanation may also be the degree of thermalization varying from
one environment to another as explained in \citet{Nara08}.

In the same way, it is noticeable that a small discrepancy between
the linear regressions and the observations appears for sources
showing the highest infrared luminosities (see Fig.
\ref{fig:corre} in the upper right corners). However, contrarily
to the previous case (nearby sources), this effect is not seen for
the SFR-CO(7-6) plot nor for the other SFR-higher-J CO
correlations. On the contrary, it appears clearer for lower-J CO
lines. One very simple first assumption should be that this
discrepancy may be caused by the fact that such active (high-z)
galaxies is expected to have a higher percent of dense gas than in
local galaxies. Indeed these objects are known to undergo more
violent star-burst phases. The gas component traced by the CO(1-0)
line may not be (relatively) as abundant and as bright as in the
local Universe. This hypothesis may be confirmed by the
localization of the CO SED maximum seen rather at high-J CO lines
(see Fig. \ref{fig:CO_curves}) which are known to trace a warmer
and denser gas (see \citealt{Baye04, Baye06}).

\subsection{Comparison with model predictions}\label{subsec:comp}

\begin{figure}
    \centering
    \includegraphics[width=8cm]{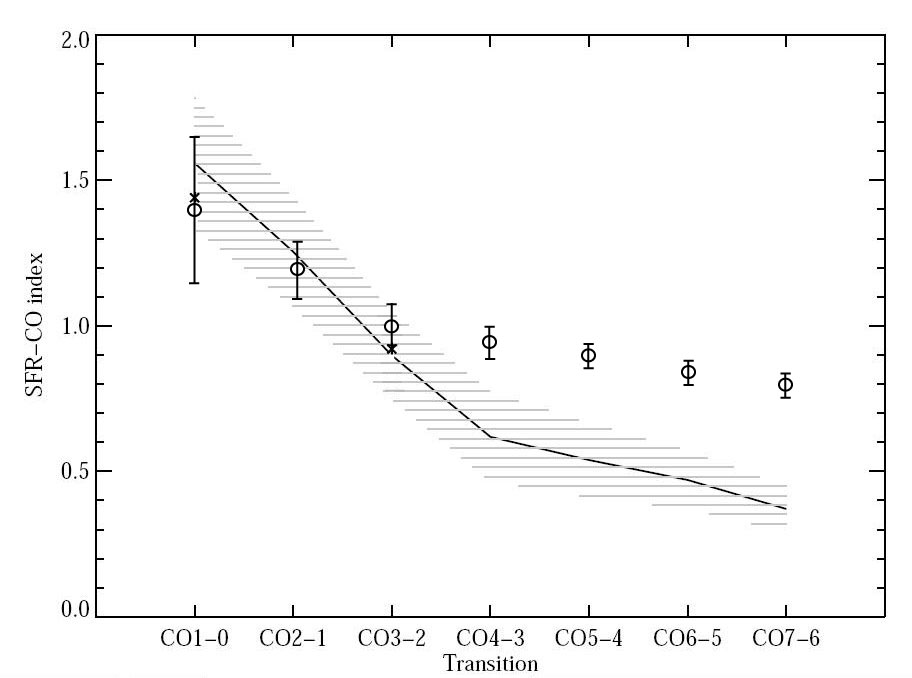}
    \caption{Predicted slopes in log(L$_{IR}$)-log(L'$_{CO}$) space as a function of
    J transitions of CO, as seen in the study of
    \citet{Nara08} (Figure 7). The grey horizontal lines
    are the model predictions and the crosses, some
    observations \citet{Nara08} used. Observational
    constraints on the slope measured from CO(1-0) to CO(7-6)
    presented in this paper are symbolized by open white circles
    (with error bars). From the CO(4-3) line, the best-fit
    slopes (dashed line) given by the predictions from \citet{Nara08} become not consistent
    with our observational results. }\label{fig:mod}
\end{figure}

It is essential to confront the results we have obtained to the
model developed by \citet{Krum07, Nara08}. More especially,
\citet{Nara08} provided some estimations of linear regression
slopes for various SFR-CO line luminosity relationships. In Fig.
\ref{fig:mod}, we have reproduced the Figure 7 of \citet{Nara08}
adding the new observational constraints we have obtained (see the
open white circles with error bars in Fig. \ref{fig:mod}). Fig.
\ref{fig:mod} thus provides the variation of the slope of the
SFR-CO luminosity relationship with respect to the transition of
CO investigated. The best predicted (from models) SFR-CO slopes
are represented by a dashed black line while the horizontal grey
lines symbolize its scatter. This scatter is computed by randomly
drawing a sample of 19 galaxies (which is similar to the size of
our nearby and high-z source sample) out of a set of 100 model
galaxies 1000 times. Here, our measurements are consistent with
other measurements (see the crosses in the Fig. \ref{fig:mod}) and
with model predictions up to the SFR-CO(3-2) relationship.
However, for higher-J CO line, some discrepancies are seen. As
mentioned in \citet{Nara08}, these discrepancies may likely due to
the fact that these models do not include any 
high-z sources. Most of these sources have an increasing
fraction of AGN as compared with this of the local galaxies. These
embedded AGN might thus significantly contaminate the infrared 
luminosity of the sources \citep{Tran01, Kim02, Veil02}, as well as be responsible of an 
increase of the gas temperature, leading to a higher slope in the 
linear regression. In a simple view, the star-formation being warmer 
in such distant objects, the temperature effect is expected especially enhanced 
for the high-J CO line relationships. Temperature contamination, 
imperfectly taken into account in the models, is thus more expected 
in such CO lines than in low-J CO transitions.  

\begin{table}
\caption{Results of the linear regressions (slope and correlation
coefficients) for the relationships between SFR and the CO
transitions from J=1-0 to J=12-11 (see Fig. \ref{fig:corre} for
some examples of SFR-CO line correlations). The linear regressions
have been obtained using the software xmgrace, including the error
bars of the observations in the calculations. In these linear
regressions, we excluded two sources from the calculations: IC~10
for all the regressions and NGC~7817 for the SFR-CO(3-2)
relationship (see Sect. \ref{subsec:gene}). We have isolated the
SFR- total CO luminosity relationship from others because we have
used different units.}\label{tab:resu_regres}
\begin{center}
\begin{tabular}{ccccc}
\hline
Molecular line & Slope & Correlation & Number\\
Luminosities &  & Coeff. & of sources\\
\hline
CO(1-0)$^{a}$ & 1.41$\pm$0.26 & 0.82 & 17\\
CO(1-0)$^{b}$ & 1.14$\pm$0.08 & 0.82 & 103\\
CO(2-1) & 1.20$\pm$0.10 & 0.95 & 17\\
CO(3-2)$^{a}$ & 1.00$\pm$0.07 & 0.97 & 17\\
CO(3-2)$^{b}$ & 0.99$\pm$0.06 & 0.89 & 81\\
CO(4-3) & 0.94$\pm$0.05 & 0.98 & 17\\
CO(5-4) & 0.90$\pm$0.04 & 0.98 & 17\\
CO(6-5) & 0.86$\pm$0.04 & 0.98 & 17\\
CO(7-6) & 0.80$\pm$0.04 & 0.98 & 17\\
CO(8-7) & 0.74$\pm$0.05 & 0.97 & 17\\
CO(9-8) & 0.67$\pm$0.06 & 0.94 & 17\\
CO(10-9) & 0.61$\pm$0.07 & 0.92 & 17\\
CO(11-10) & 0.57$\pm$0.07& 0.90 & 17\\
CO(12-11) & 0.53$\pm$0.07& 0.89 & 17\\
\hline
CO(tot) & 0.62$\pm$0.04 & 0.97 & 16$^{c}$\\
\hline
\end{tabular}
\end{center}

$^{a}$: Without any additional literature data; $^{b}$: With
additional literature data; $^{c}$: Markarian 231 is not included
in this linear regression calculation because we did not find any
estimation of its CO(11-10) and CO(12-11) lines luminosities
needed for estimating properly its total CO luminosity.
\end{table}

\section{Conclusions}\label{sec:conclu}

In this paper we have investigated the relevance of using the CO
line emissions as indicators of star formation rate. For the first
time, we have both studied this question over a large number of CO
transitions (12) and over a large sample of source going up to a
redshift of z$>$6. We have shown that the SFR-total CO,
-CO(5-4),-CO(6-5) and -CO(7-6) line luminosity relationships are
the tightest correlations, making these lines the best indicators
of the star formation rates. The results we have obtained also
strongly confirmed the predictions from \citet{Krum07, Nara08} who
showed that the SFR-CO lines luminosity relationships are above
all regulated by the Kennicutt-Schmidt law, which sets the way in
which observed transitions trace molecular gas. In other words,
the SFR-CO line luminosity relationships depends on how the
critical density of the molecule compares to the mean density of
observed source. We confirm in this paper that, for the CO lines
with J$_{upper}>$3, the SFR-CO line luminosity relationship are
indeed sub-linear. However some discrepancies between model
predictions and observations appear for higher-J CO lines (from
CO(4-3)) relationships. These differences may be likely due to the
fact that, in the model the high-z sources are not included
(possible temperature contamination from AGN heating processes). Anyway, more
observations (ALMA, Herschel) of CO in both nearby and high-z
sources, especially the high frequency CO transitions could also
be very helpful for explaining these discrepancies.

\section*{Acknowledgments}

EB acknowledges financial support from the Leverhulme Trust.

\newcommand{\apj}[1]{ApJ, }
\newcommand{\aj}[1]{Aj, }
\newcommand{\apjs}[1]{ApJS, }
\newcommand{\apjl}[1]{ApJ Letter, }
\newcommand{\aap}[1]{A\&A, }
\newcommand{\aaps}[1]{A\&A Suppl. Series, }
\newcommand{\araa}[1]{Annu. Rev. A\&A, }
\newcommand{\aaas}[1]{A\&AS, }
\newcommand{\bain}[1]{Bul. of the Astron. Inst. of the Netherland,}
\newcommand{\mnras}[1]{MNRAS, }
\newcommand{\araaa}[1]{ARA\&A, }
\newcommand{\planss}[1]{Planet Space Sci., }
\newcommand{\jrasc}[1]{Jr\&sci, }
\newcommand{\pasj}[1]{PASJ, }

\bibliographystyle{mn2e}
\bibliography{references}

\appendix \label{app:LVG}

\section{Single component LVG modelling results for the high-z sources}

As mentioned in Subsect. \ref{subsec:highz}, a complete
\emph{observed} CO SED does not exist for J=1-0 to J=12-11
in high-z sources. We thus have collected CO line emissions from
\citet{Solo05} and performed a Large Velocity Gradient (LVG)
analysis to derive the missing CO integrated line intensities.
The results of this modelling work is shown in Fig. \ref{fig:CO_curves}
and in the following Table \ref{tab:lvg}.

We have run LVG models \citep{Gold74, DeJo75}, already well
described in various papers. The version we have used in this
study is the one presented in \citet{Baye04, Baye06, Baye08b}.
This LVG model is basically running with three free parameters:
the gas density n(H$_{2}$), the kinetic temperature (T$_{k}$) and
the CO column density divided by the line width (N($^{12}$CO)$/
\Delta v$). We have investigated the following range of LVG input
parameters : 5 K $<$T$_{K}<$ 255 K, $1 \times
10^{12}$cm$^{-2}/$kms$^{-1}$$<$N($^{12}$CO)$/ \Delta v < 1 \times
10^{19}$cm$^{-2}$ and $1 \times 10^{1}$cm$^{-3}<$n(H$_{2}$)$<1
\times 10^{7}$cm$^{-3}$.

For determining the missing CO line intensities, we have
constrained the predicted CO line intensity ratios with the
observed values (computed from \citealt{Solo05}) via a reduced
$\chi^{2}$ method as performed in \citet{Baye04, Baye06, Baye08b}.

We have used a single LVG component for reproducing the entire set
of CO data in high-z sources, being fully conscious that this
modelling approach is not ideal for reproducing the widespread
view that the ISM contains dense, star-forming molecular cloud
cores and more diffuse gas in cloud envelopes, even in high-z
sources. Motivated also by, on average, the small number of CO
detections, we have thus considered this approach as reasonable
for the purpose of this paper.

We recommend to use the predicted physical properties derived and
listed in Table \ref{tab:lvg} as only indicative values. We indeed
remind the reader that we were not trying to model each source
individually, deriving the accurate set of physical properties for
the molecular gas, but that we rather aim to obtain satisfactory
estimations of the CO line intensities in these sources. Note
that, for APM 08279 \citep{Weis07}, the use of two LVG components
has been made but, such as the sum of \textit{both} CO components
agrees with the observed CO SED. Thus, the choice of a one or a
two LVG components model does not affect significantly the
SFR-\textit{total} CO luminosity relationships we present in this
paper.

\begin{table}
\caption{Results of the single component LVG model analysis. We
presented in this table the physical properties of the best LVG
model (having the lowest-$\chi^{2}$ value when compared to the
observations).}\label{tab:lvg}
\begin{tabular}{ccccc}
\hline
Source & T$_{K}$ & n(H$_{2}$) & N(CO)$/ \Delta v$ \\
 & in K &  ( cm$^{-3}$) & (cm$^{-2}$/kms$^{-1}$) &\\
\hline
HR10        & 255 & 1.0$\times 10^{1}$ & 8.0$\times 10^{18}$\\
IRAS F10214 & 15 & 8.0$\times 10^{5}$ & 3.0$\times 10^{18}$\\
Cloverleaf  & 20 & 4.0$\times 10^{4}$ & 1.0$\times 10^{19}$\\
SMM J14011  & 205 & 2.7$\times 10^{1}$ & 7.9$\times 10^{18}$\\
VCV J1409   & 185 & 1.7$\times 10^{4}$ & 4.5$\times 10^{15}$\\
4CO60.07    & 180 & 1.7$\times 10^{2}$ & 1.5$\times 10^{18}$\\
APM 08279   & 255 & 9.0$\times 10^{3}$ & 8.5$\times 10^{17}$\\
PSS J2322   & 30 & 5.0$\times 10^{4}$ & 2.2$\times 10^{16}$\\
TN J0924    & 205 & 1.4$\times 10^{2}$ & 2.5$\times 10^{18}$\\
SDSS J1148  & 100 & 7.9$\times 10^{4}$ & 4.5$\times 10^{15}$\\
\hline
\end{tabular}
\end{table}

\bsp

\label{lastpage}

\end{document}